\documentclass[]{spie}  

\usepackage{amsmath,amsfonts,amssymb}
\usepackage{graphicx}
\usepackage[dvipsnames]{xcolor}
\usepackage[colorlinks=true, allcolors=RoyalBlue]{hyperref}
\usepackage{siunitx}
\usepackage[]{xspace}
\usepackage{enumitem}
\usepackage{perpage} 
\MakePerPage{footnote}
\usepackage [english]{babel}
\usepackage [autostyle, english = american]{csquotes}

\usepackage[table]{xcolor}
\usepackage{multirow}
\usepackage{tabularx}
\usepackage{makecell}
\usepackage{array}

\definecolor{BandBlue}{RGB}{220,235,250}
\definecolor{BandOrange}{RGB}{255,228,196}
\definecolor{BandRed}{RGB}{250,215,215}
\definecolor{SupportGreen}{RGB}{220,245,220}
\definecolor{BandGrey}{RGB}{245,245,245}

\def\bi{\begin{itemize} \itemsep1pt \parskip2pt \parsep0pt}
\def\ei{\end{itemize}}

\def\arcsec{\hbox{$^{\prime\prime}$}}

\newcommand{\jwst}{{\bf \href{https://science.nasa.gov/mission/webb/}{jwst}}\xspace}
\newcommand{\rst}{{\bf \href{https://roman.gsfc.nasa.gov}{Roman}}\xspace}
\newcommand{\hwo}{{\bf \href{https://habitableworldsobservatory.org/}{HWO}}\xspace}

\newcommand{\cpp}{{\bf \href{https://www.romancoronagraph.space}{CPP}}\xspace}
\newcommand{\cgi}{{\bf \href{https://science.nasa.gov/mission/roman-space-telescope/coronagraph/}{Coronagraph Instrument}}\xspace}
\newcommand{\cg}{{\bf \href{https://science.nasa.gov/mission/roman-space-telescope/coronagraph/}{Coronagraph}}\xspace}
\newcommand{\cgisim}{{\tt \href{https://sourceforge.net/projects/cgisim/}{CGISim}}\xspace}
\newcommand{\corgisim}{{\tt \href{https://corgisim.readthedocs.io/en/latest/}{CorgiSim}}\xspace}

\newcommand{\corgi}{{\tt \href{https://github.com/roman-corgi/}{roman-corgi}}\xspace}
\newcommand{\corgidrp}{{\tt \href{https://corgidrp.readthedocs.io}{corgidrp}}\xspace}

\newcommand{\nircam}{{\bf \href{https://jwst.stsci.edu/instrumentation/nircam}{NIRCam}}\xspace}

\newcommand{\orbitize}{{\tt \href{https://orbitize.readthedocs.io/}{orbitize!}}}

\newcommand{\arcdeg}{\mbox{$^{\circ}$}}

\newcommand{\mjup}{\mbox{ $\mathrm{M_{Jup}}$}\xspace}

\def\begini{\begin{itemize}[itemsep=0.5pt,topsep=0.5pt]}
\def\endi{\end{itemize}}

\usepackage{todonotes}

\title{The Roman Coronagraph Community Participation Program: early calibration plan and pilot observation of a companion}

\author[a]{Julien H. Girard} 
\author[b]{Eric Cady} 
\author[c]{Neil T. Zimmerman} 
\author[d]{Clarissa R. Do Ó} 
\author[e]{Jingwen Zhang} 
\author[f]{Guillermo Gonzalez}
\author[f]{Bijan Nemati}
\author[b]{Vanessa P. Bailey} 
\author[g]{Alexis Lau} 
\author[g]{Sophie Noiret} 
\author[b]{John Krist}
\author[b]{Julia Milton} 
\author[b]{Marie Ygouf} 
\author[h]{Ramya M. Anche}
\author[h]{Schuyler Wolff} 
\author[h]{Justin Hom} 
\author[i]{Amanda Chavez} 
\author[i]{Jason J. Wang} 
\author[e]{Maxwell Millar-Blanchaer} 
\author[j]{Jessica Gersh-Range} 
\author[k]{Matthias Samland} 
\author[k]{Macarena Vega-Pallauta} 
\author[k]{Wolfgang Brandner} 
\author[l,m]{Toshiyuki Mizuki} 
\author[l,m]{Masayuki Kuzuhara} 
\author[b]{Susan Redmond} 
\author[n]{Kevin Ludwick} 
\author[e]{Jaren N. Ashcraft} 
\author[o]{William O. Balmer} 
\author[p]{Rusland Belikov} 
\author[q]{Beth Biller} 
\author[r]{Sarah Blunt} 
\author[g]{Ellis Bogat} 
\author[k]{Óscar Carrión-González} 
\author[k]{Gael Chauvin} 
\author[s]{Alexandra Z. Greenbaum} 
\author[t]{Sebastiaan Haffert} 
\author[s]{Jim Ingalls} 
\author[b]{Brian Kern}
\author[b]{Bertrand Mennesson} 
\author[a]{Laurent Pueyo} 
\author[u]{Dmitry Savransky} 
\author[p]{Dan Sirbu} 
\author[f]{Peter Williams} 
\author[o]{Chen Xie} 

\affil[a]{\small Space Telescope Science Institute, 3700 San Martin Dr, Baltimore MD, 21218, USA}
\affil[b]{\small Jet Propulsion Laboratory, California Institute of Technology, Pasadena, CA 91109, USA}
\affil[c]{\small NASA Goddard Space Flight Center, Greenbelt, MD 20771, USA}
\affil[d]{\small Department of Astronomy, California Institute of Technology, Pasadena, CA 91125, USA}
\affil[e]{\small Department of Physics, University of California, Santa Barbara, CA 93106, USA}
\affil[f]{\small Tellus1 Scientific, LLC, 8401 Whitesburg Dr SE, Unit 4662, Huntsville, AL 35802 USA}
\affil[g]{\small Aix Marseille Univ, CNRS, CNES, LAM, Marseille, France}
\affil[h]{\small Steward Observatory \& Department of Astronomy, Univ of Arizona, 933 N Cherry Avenue, Tucson AZ 85721, USA}
\affil[i]{\small CIERA and Department of Physics \& Astronomy, Northwestern University, Evanston, IL 60208, USA}
\affil[j]{\small DM Telescopes LLC, Raleigh, NC, USA}
\affil[k]{\small Max Planck Institute for Astronomy, Konigstuhl 17, 69117 Heidelberg, Germany}
\affil[l]{\small Astrobiology Center of NINS, 2-21-1, Osawa, Mitaka, Tokyo, 181-8588, Japan}
\affil[m]{\small National Astronomical Observatory of Japan, 2-21-2, Osawa, Mitaka, Tokyo, 181-8588, Japan}
\affil[n]{The University of Alabama in Huntsville, 301 Sparkman Drive, Huntsville, AL 35899, USA}
\affil[o]{\small Department of Physics \& Astronomy, Johns Hopkins University, 3400 N. Charles St., Baltimore, MD 21218, USA}
\affil[p]{\small NASA Ames Research Center, Moffett Field, CA 94035, USA}
\affil[q]{\small Institute for Astronomy, Centre for Exoplanet Science, University of Edinburgh, Edinburgh EH9 3HJ, UK}
\affil[r]{\small Department of Astronomy and Astrophysics, University of California, Santa Cruz, Santa Cruz, CA 95064, USA}
\affil[s]{\small NASA Exoplanet Science Institute, IPAC, California Institute of Technology, Pasadena, CA 91125 USA}
\affil[t]{\small Sterrewacht Leiden, PO Box 9513, Niels Bohrweg 2, Leiden, The Netherlands}
\affil[u]{\small Sibley School of Mechanical and Aerospace Engineering, Cornell University, Ithaca, NY, 14853, USA}

\authorinfo{Further author information: send correspondence to Julien Girard ({\tt \href{mailto:jgirard@stsci.edu}{jgirard@stsci.edu}})}
 
\begin{document} 
\maketitle

\vspace{-0.17cm}
\begin{abstract}
The Nancy Grace Roman Space Telescope is set to launch on August 30, 2026 (and "no later than the Spring 2027"), a month away at the time of submission of this paper!  The Coronagraph Instrument - technology pathfinder for future direct imaging missions - is ready to fly onboard Roman. According to predictions, laboratory tests and high fidelity simulations, it will open a new contrast regime that enables the imaging of mature, giant planets in visible reflected light. The Community Participation Program is responsible for preparing a comprehensive observing program for the Coronagraph with associated data processing software and calibrations. In this paper we give a brief update about the on-going "baseline" calibration plan for the first months. Additionally, we describe the preparation of a pilot program with the observation of the stellar companion HD 29992 B at moderate ($\sim10^{-5}$ to $\sim10^{-6}$) Band 1 (575 nm) contrast. This program, focusing on the only required (fully supported) mode is to be carried out as soon as the instrument is operational. The idea is to generate a canonical data set of a bright star with a self luminous companion that is easily recoverable by the Data Reduction Pipeline. This functional checkout and “training” data set will be precious to best prepare our community, exercise our calibration plan on an "easy" companion to then tackle more challenging observing campaigns.
\end{abstract}

\keywords{High Contrast Imaging, Coronagraphs, Space Telescopes, Calibration, Exoplanets, Reflected Light}

\section{A new era for space coronagraphy}
\label{sec:intro}  

\begin{figure}[h!]
\begin{center}
\includegraphics[width=0.7\textwidth, angle=0]{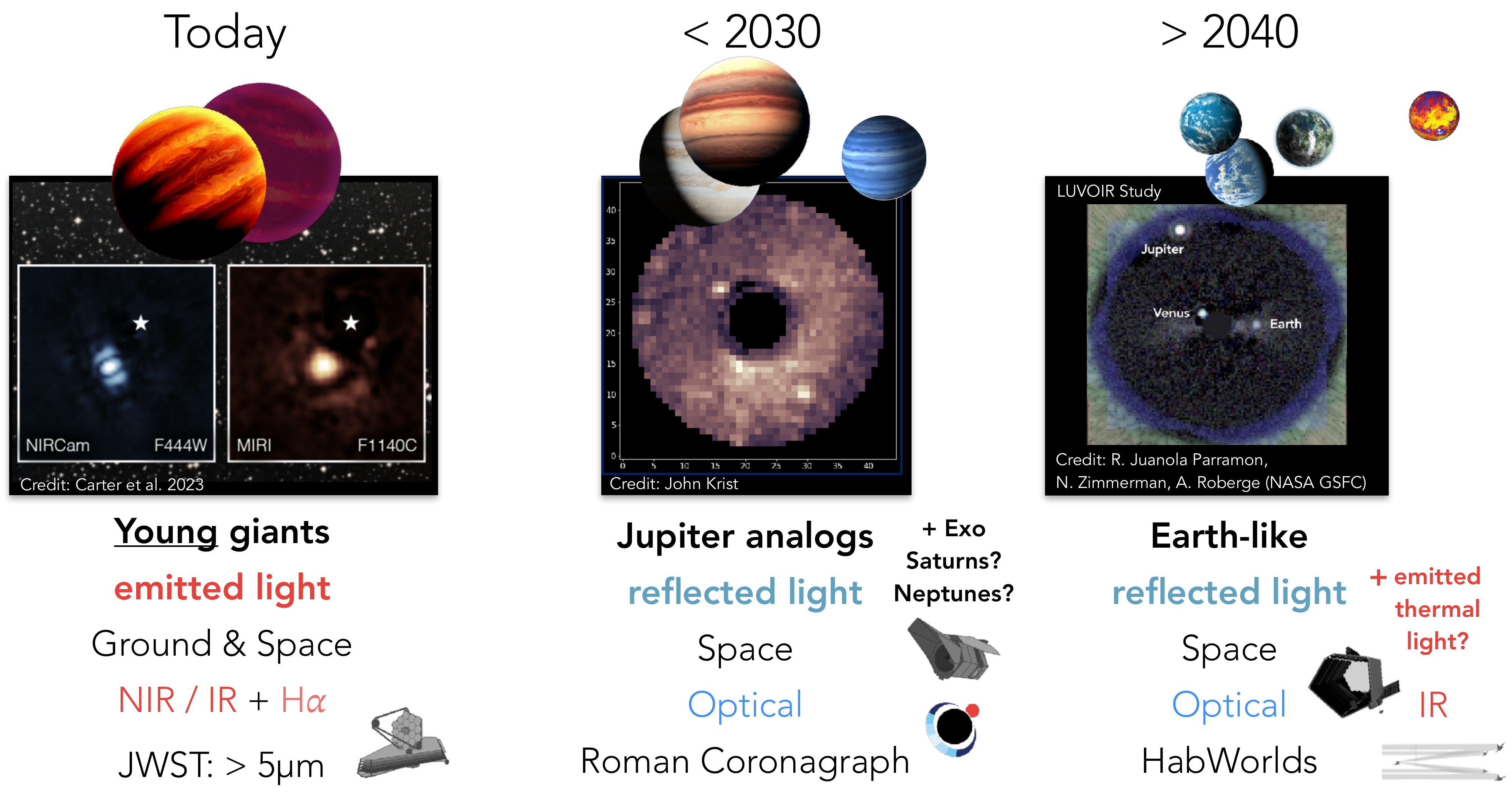}
\caption{Direct Imaging "roadmap" from images of young giant exoplanets Today (real images)\cite{carter2023} to the first images of giant planets in reflected light with the \cgi (simulations) hopefully before 2030 to future, post 2040 images of Earth analogs and rocky planets with missions like HWO\cite{roberge2018luvoir} (reflected light direct imaging) and perhaps LIFE\cite{quanz2022} (nulling interferometry in thermal imaging) which do present advantageous synergies\cite{carrion-gonzalez2023}.}
\label{fig:roadmap}
\end{center}
\end{figure}

\noindent As of Today we are able to directly imaged young, dusty and still hot (a few hundreds to $\sim$1,800 K in effective temperature) giant exoplanets around nearby stars\cite{bowler2016_review, currie2023}. JWST coronagraphs\cite{boccaletti2022, girard2022} have started to extend the direct imaging and spectroscopic campaigns to more mature, colder giant planets\cite{matthews2024, bardalez2025} and/or less massive young planets down to about a Saturn mass for the most extreme case to date\cite{lagrange2025, crotts2025}. \noindent Finding through direct imaging and studying potentially habitable Earth-like planets around Sun-like stars is a dream our high contrast imaging community has had since decades\cite{roberge2018luvoir}. A telescope built with that explicit goal is under study, the recently announced NASA Habitable Worlds Observatory (\hwo)\cite{feinberg2024, stark2024, feinberg2026}. To detect such exoplanets in visible reflected light, one needs to achieve contrast ratios over 10 billions. This exciting future evolution of the exoplanet imaging endeavor is summarized in figure~\ref{fig:roadmap} with illustrations and data (real for \jwst, simulated for the \rst \cgi and \hwo (LUVOIR concept study in this case).

\vspace{-0.5cm}
\subsection{The Roman Coronagraph technology demonstration instrument }
\label{sec:cgi}  

\noindent The \rst Space Telescope (\rst hereafter, formerly the Wide Field Infrared Survey Telescope or WFIRST) has a primary mirror that is 2.4 meters (7.9 feet) across, which is the exact same size as that of the Hubble Space Telescope. But at f/7.9 (HST is f/24), \rst is optically optimized for wide-field astronomy in the near-infrared and most of its lifetime will be spent surveying large portions of the sky with its prime, Wide Field Instrument (WFI)\cite{schlieder2024}. 
 
\noindent The \cgi\cite{mennesson2022}(previously and informally called "CGI" amongst stakeholders) aboard the Nancy Grace \rst Space Telescope will demonstrate critical technologies\cite{baker2025, kuan2025, zhou2025},  that will pave the way for the Habitable Worlds Observatory\cite{feinberg2026}. Current predictions place the Roman Coronagraph's detection limit between approximately $\sim 10^{-8}$ and a few $\times 10^{-9}$, enabling groundbreaking science such as the direct imaging of Jupiter-like exoplanets orbiting mature stars and the detection of exozodiacal dust disks in scattered light. 

\noindent But first, the \cgi will have to pass the so called "Technology Demonstration Requirement at Level 1" or TTR5\cite{poberezhskiy2025, cady2025} which is that  "Roman shall be able to measure (using its \cg), with a signal to noise ratio (SNR) $\leqslant$ 5, the brightness of an astrophysical point source located between 6 and 9 $\lambda/D$ from an adjacent star with a $V_{AB}$ magnitude $\leqslant$ 5, with a flux ratio $\leqslant 10^{-7}$; the bandpass shall have a central wavelength $\leqslant$ 600 nm and a bandwidth $\geqslant$ 10\%.  

\begin{figure}[h!]
\begin{center}
\includegraphics[width=0.7\textwidth, angle=0]{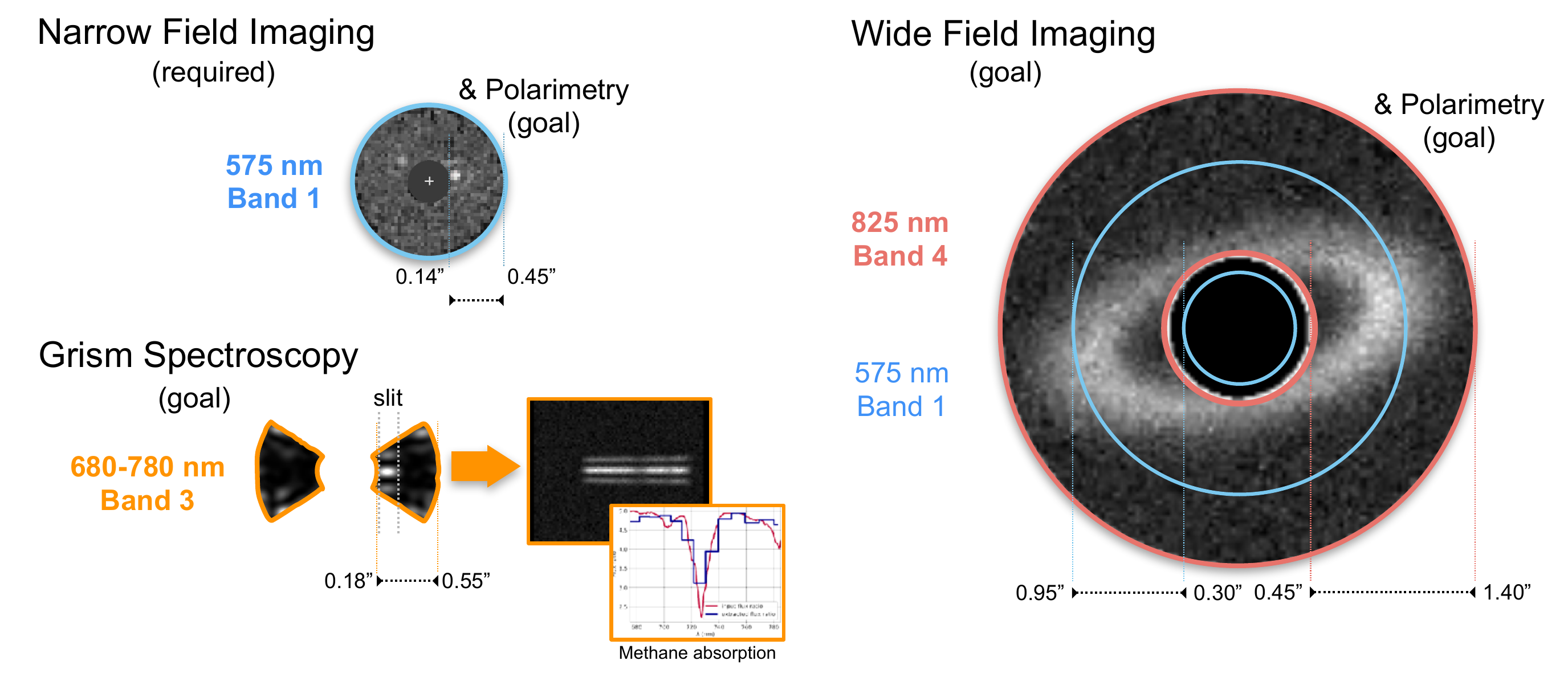}
\vspace{0.1cm}
\caption{The main \cgi modes (also listed in table~\ref{tab:modes} and working angles. The images and spectrum are simulations made for illustration purposes only.}
\label{fig:modes}
\end{center}
\end{figure}

\noindent To achieve such contrasts despite a complex entrance aperture with a large central obscuration, the \cg is using quite complex and  Lyot stops and its throughput is a few percents at most. This has an implication that the whole technology demonstration and even the observation phase has to be done using bright stars. Digging the dark hole on stars of V$\geqslant$3 would be prohibitively long. Despite numerous limitations (i.e the so called "ground-in-the-loop" wavefront control scheme which induces a factor two hit in efficiency), the \rst \cg, with its main modes (summarized in figure~\ref{fig:modes} as well as in the "Primer" document on the \cpp website) is extremely transformative, on time,  and an exciting pathfinder for \hwo.

\noindent In Table~\ref{tab:modes} we list the main modes of the \cg including the only fully supported (because required for the technology demonstration) Hybrid Lyot Coronagraph Narrow Field Band 1 (HLC NF B1) with an inner-working angle (IWA) of $\sim$0.14\arcsec. All other modes are so called "goal modes" that will be commissioned on "Best Effort" basis. There are many other modes and mask / filter / field-stop combinations\cite{riggs2025} that could potentially be tested further down the road but at this point the CTC, SSC, SOC and \cpp have to compose with the 18-month observing program, no mission extension is being discussed. For this reason, the initial, on-sky "baseline" calibration plan summarized in section~\ref{sec:calplan} focuses primarily on the HLC NF B1 mode, that of the TTR5 demonstration.

\begin{table}[h!]
\caption{Required and Best Effort Modes for the Roman Coronagraph Instrument.}
\label{tab:modes}
\begin{center}
\small
\renewcommand{\arraystretch}{1.1}
\begin{tabular}{|c|c|c|c|c|c|}
\hline
\cellcolor{BandGrey}\textbf{Band} &
\cellcolor{BandGrey}\textbf{Mode} &
\cellcolor{BandGrey}\textbf{Mask Type} &
\cellcolor{BandGrey}\textbf{FoV} &
\cellcolor{BandGrey}\textbf{Coverage} &
\cellcolor{BandGrey}\textbf{Support} \\
\hline

\cellcolor{BandBlue}\shortstack{\textbf{1}}

&
Narrow FoV
&
Hybrid Lyot
&
$0.^{\prime\prime}14-0.^{\prime\prime}45$
&
360$^\circ$
&
\cellcolor{SupportGreen}Imaging: Required
\\
\cellcolor{BandBlue}\shortstack{575 nm}
&
Imaging (HLC NF B1)
&
&
&
&
{\small Polarimetry: Best Effort}
\\
\hline
\cellcolor{BandBlue}\shortstack{\textbf{1}}
&
Wide FoV
&
Shaped Pupil
&
$0.^{\prime\prime}30-0.^{\prime\prime}95$
&
360$^\circ$
&
Imaging: Best Effort
\\
\cellcolor{BandBlue}\shortstack{575 nm}
&
Imaging: Best Effort
&
&
&
&
{\small Polarimetry: Best Effort}
\\
\hline

\cellcolor{BandOrange}\shortstack{\textbf{3}}
&
Slit + $R\sim50$
&
Shaped Pupil
&
$0.^{\prime\prime}18-0.^{\prime\prime}55$
&
2$\times$65$^\circ$
&
{\small Spectroscopy: Best Effort}
\\
\cellcolor{BandOrange}\shortstack{730 nm}
&
Prism Spectroscopy
&
&
&
&
\\
\hline

\cellcolor{BandRed}\shortstack{\textbf{4}}
&
Wide FoV
&
Shaped Pupil
&
$0.^{\prime\prime}45-1.^{\prime\prime}4$
&
360$^\circ$
&
Imaging: Best Effort
\\
\cellcolor{BandRed}\shortstack{825 nm}
&
Imaging
&
&
&
&
Polarimetry: Best Effort
\\
\hline

\end{tabular}
\end{center}
\end{table}

\vspace{-0.3cm}
\subsection{The Community Participation Program (CPP)}
\label{sec:cpp}  

\noindent The \cpp\cite{wolff2026tmp} was established through the NASA ROSES ROMAN-22 program to maximize the scientific and technological return of the \cgi while preparing the broader exoplanet imaging community to fully exploit its unprecedented capabilities. Working in close collaboration with the Roman Project, Instrument, and Science Support Center (SSC at IPAC) and Science Operations Center (SOC at STScI) teams, the CPP is responsible for developing the infrastructure needed to plan, execute, and analyze Coronagraph observations. These efforts include the development of the \texttt{CorgiETC} exposure time calculator, end-to-end observation simulation tools (\cgisim\cite{krist2023}and its wrapper \corgisim\cite{zhang2026tmp}, target databases\cite{savransky2024, savransky2026tmp}, observation planning softwares (public and private \corgi GitHub repositories), and a community data reduction and analysis pipeline (DRP)\cite{wang2026tmp, millarblanchaer2024}. In parallel, the \cpp coordinates preparatory observations, target selection, reference star vetting\cite{hom2026,hom2026tmp}, community white paper calls, documentation, tutorials, and workshops to ensure that both the technology demonstration objectives and the highest-priority exoplanet science cases are realized. Collectively, these software tools and community resources provide a common framework for designing observations, validating analysis techniques, interpreting high-contrast imaging data, and reducing technical risk for future flagship missions such as \hwo. The \cpp capitalizes on previous work from two Science Investigation Teams (SIT) prior to 2022\cite{kasdin2020, girard2020, turnbull2021, bailey2023, savransky2024} and has grown a lot in the past two years with notably many early career scientists.

\subsection{Timeline, transition from commissioning to science verification and operations}
\label{sec:timeline}  

\begin{figure}[h!]
\begin{center}
\includegraphics[width=1.0\textwidth, angle=0]{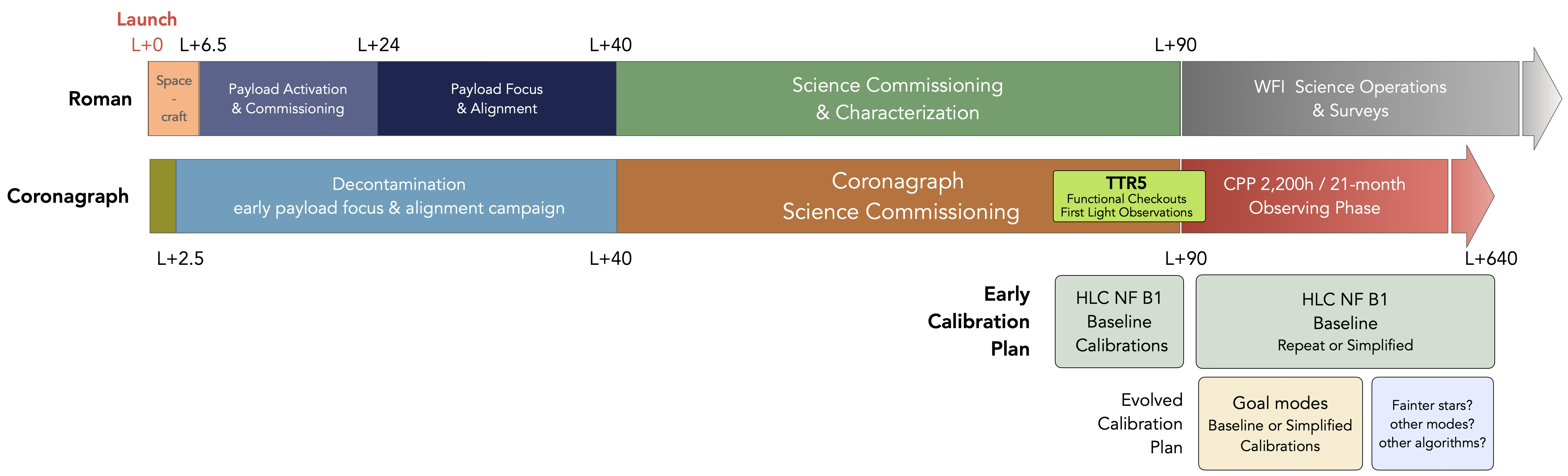}
\vspace{0.2cm}
\caption{Timeline (adapted from Eric Cady's Commissioning Plan presentation at Spirit of Lyot 6\cite{cady2026}) from the launch (L+0 days) of \rst to the end of the 21-month \cpp-led observation phase.}
\label{fig:timeline}
\end{center}
\end{figure}

\noindent Figure~\ref{fig:timeline} shows the time for both \rst in general and the \cg starting from launch and using units of days. Launch (L+0) is currently set on August 30, 2026 at 11:26 UTC (07:26 ETC). Most of the \cpp work start at L+90 but the TTR5 and hopefully the full, "baseline" (Early) calibration plan with the HLC NF B1 mode will be carried out and tested as soon as the instrument is able to dig and sustain a dark hole, at the end of the commissioning. The "goal" modes (wide field imaging, polarimetry, spectroscopy, etc.) will be tested and calibrated during the 2,200 Observing Phase.

\vspace{-0.3cm}
\section{On-sky Early Calibration Plan}
\label{sec:calplan}  

\noindent The \cgi aims at demonstrating technologies but also learning as much as possible on science operations with wavefront control in space fundamental contrast limits with the given hardware, algorithms, observations strategies. Of course, the \cpp also designed the program to attempt important science goals, most notably on the detection and characterization of giant exoplanets in visible reflected light for the firs time. It will be a formidable opportunity to learn about calibration as well, our capacity to operate for many tens of hours in a new contrast regime and in photon counting mode, digging and maintaining a dark hole, being able to recover point sources and diffused exozodiacal light with various PSF subtraction strategies, etc. The Electron-Multiplying Charge Coupled Device (EMCC) detector alone is prone to degradation and requires a set of specific calibration activities. 

\subsection{Brief Update: Focus on Completeness}
\label{sec:calstrategy}  

\noindent In Zellem et al. 2022\cite{zellem2022}, an initial Calibration Plan was defined using current best estimates (CBE) on Flux Ratio Noise (FRN) uncertainties derived for each calibration product based on error budgets and simulations. For each calibration a quite large margin was applied. With Today's more accurate knowledge on how to estimate the clock time necessary to carry out each activity, we have decided to keep all calibration items because they are all important and we will learn what limits our flux ratio measurements once on sky. Nevertheless, the strategy has evolved a little bit towards maximizing the time spent attempting challenging reflected light planets. This is especially true at the very beginning of the \cpp's Observing Phase. We think we will learn a lot from performing shallower calibration observations (i.e sparser CT pattern, only 3 faint CALSPEC stars instead of 10 in Zellem et al. 2022) but check the repeatability of the products with a once/month or once/campaign (every time there is a mode change or a new dark hole is initiated) as well as extend it to "goal" modes.

\noindent To calibrate first look observations and the TTR5 (i.e. calibrate the contrast), we absolutely want to have at least one good set of all baseline calibrations as early as possible, hopefully during the end of the 90-day commissioning period (see. figure~\ref{fig:timeline}). It is crucial to be able to calibrate all data and make sure the data reduction pipeline and analysis software functions properly with the correct headers, producing the correct product for other pipeline blocks, etc.  We think this "completeness first" and agile approach will enable the team and community to learn more and mitigate the risk of missing reflected light planets as several top ranked targets happen to be suitable very early in the observing phase with the option to refine the calibrations if needed. Below is a list of most science calibrations (not including wavefront control and more early functional tests):
\bi
\small 
\item The {\bf Absolute Flux} calibration
measures the brightness of stars accurately so that the planet-to-star flux ratio can be determined. It is performed by observing CALSPEC standard stars\cite{bohlin2014} with and without neutral density (ND) filters and monitoring a designated "sweet spot" on the ND filter to maintain consistent measurements (of the unsaturated core). {\bf Update}: While Zellem et al. 2022\cite{zellem2022} recommended the observation of 3 bright stars and 10 faint stars to achieve a 1.41\% CBE (30\% margin with respect to the 2\% error allocation), we found that we could aching a 2.9\% CBE with only 1 bright (with ND) and 3 faint stars (without ND). In high contrast imaging, a 5\% relative photometry accuracy is often considered excellent given that the faint point source photometric extraction in the presence of speckle noise is a considerable challenge. With that we are significantly reducing the clock time for this calibration item and we are thus in capacity to repeat it more (with 1 or 3 faint stars once per campaign or once per month as explained in table~\ref{tab:calplan}). 

\begin{figure}[h!]
\begin{center}
\includegraphics[width=0.7\textwidth, angle=0]{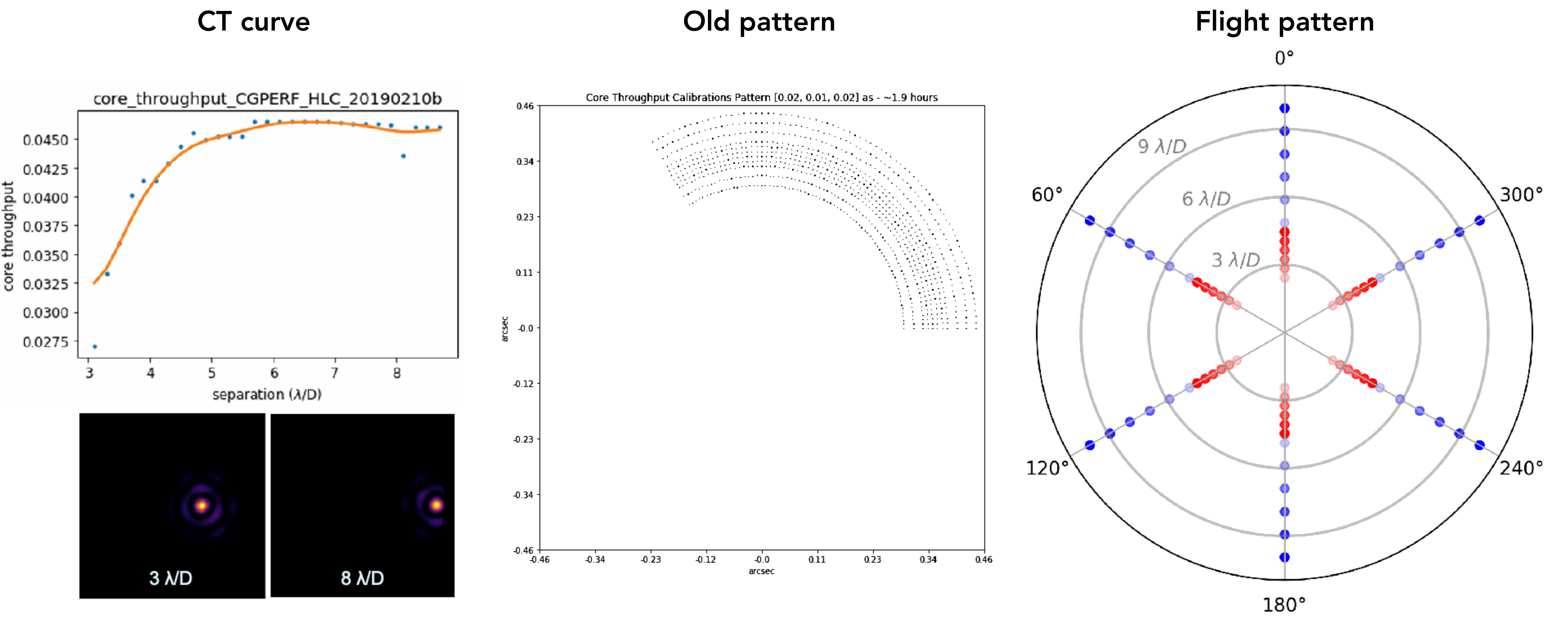}
\vspace{0.2cm}
\caption{Core Throughput calibration baseline calibration (flight) pattern  and the old pattern (used at the time of Zellem et al. 2021}. 
\label{fig:ct}
\end{center}
\end{figure}
\ei

\vspace{-1.1cm}
\bi
\item The {\bf Core Throughput (CT)} calibration measures how efficiently light from a planet is transmitted through the coronagraph system. A star is dithered across the field of view to map throughput and point spread function (PSF) variations, improving the accuracy of photometric measurements. The activity will also take a pupil image to have an additional total flux reference to be able to normalize. The many measured off-axis PSF can be use later by the DRP to forward-model "fake planets" and get accurate relative photometry. {\bf Update}: The old dithering pattern was very dense as shown in figure~\ref{fig:ct} but only covered an arc and positions from 6 to 9 $\lambda$/D (the TTR5 requirements). A lot of suitable reflected light planets will be much closer to the IWA and therefore we decided to generate off-axis PSF from 2.8 to 9 $\lambda$/D to enable CT calibration and/or forward modeling. We use a sparser but azimuthal pattern to check for eventual asymmetries. The number of positions is significantly less than before and with today's understanding of overheads, the activity still takes 2.75 hours for one mask/filter combination. Globally the throughput curve is known but we want to be able to check it on sky, check how much the CT is affected by various instances of dark holes. If we repeat it often it is important for this activity to remain under 3 hours.

\item The {\bf Astrometric} calibration determines the precise positions of objects within the telescope's field of view. It uses observations of dense star fields, such as those in the Large Magellanic Cloud\cite{chavez2026tmp}, to create distortion maps and ensure highly accurate position measurements. The products here are the orientation or "North Angle" and the plate-scale.

\item The {\bf Flat Field} calibration corrects for variations in detector sensitivity across the field of view, including large-scale, medium-scale, and pixel-to-pixel differences. Since the telescope has no internal flat-field lamp, astronomical sources are observed using dithering and raster scanning to create accurate flat-field maps. It is carried out using Neptune or Uranus (when available)\cite{maier2022}.

\item Baseline {\bf Detector} calibrations\cite{nemati2020}: The {\bf Charge Transfer Inefficiency (CTI)} calibration aims at correcting radiation-induced charge traps in the detector. Trap pumping and dark frame observations are used to identify trap locations and characteristics so that software can remove CTI effects from science images. The {\bf Cosmic Ray} calibration identifies and removes the effects of high-energy particles striking the detector. The baseline approach is to flag and mask affected pixels, although methods to recover the underlying data are also being investigated. The {\bf Darks and Clock-Induced Charge (CIC)} calibration measures detector noise caused by dark current and clock-induced charge. Many dark frames are combined to create a master dark frame, which is used to remove these background signals from science observations. The {\bf Image Correction: Nonlinearity} and {\bf K Gain} calibration activity measures how the {\tt EXCAM} detector response changes at different signal levels and determines the K gain, which converts electrons into detector counts. A photon transfer curve analysis is used to calculate detector gain, read noise, and correct non-linear pixel responses.
\item {\bf Goal Modes}\cite{groff2025}: The {\bf Polarimetry} calibration\cite{anche2026tmp, mizuki2026tmp} measures and corrects for polarization effects introduced by the telescope's optics. Observations of polarized and unpolarized standard stars are used to determine the instrument's Mueller Matrix, allowing accurate measurements of polarized light. There will also be a set of absolute flux measurements using CALSPEC standards stars to account for the Wollaston prism and field stop of this mode. The {\bf Spectroscopic} calibration establishes the wavelength scale, spectral orientation, and slit performance of the instrument. Observations of unocculted stars and narrowband filters are used to calibrate wavelength zero-points, spectral dispersion, and slit losses for accurate spectroscopy.
\ei


\begin{table*}[h!]
\centering
\scriptsize
\begin{tabular}{p{2.8cm} p{1.0cm} p{1.4cm} p{2.5cm} p{1.35cm} p{5.6cm}}
\hline 
\textbf{Calibration item} & 
\textbf{Baseline (TTR5)} & 
\textbf{Clock time (hours)} & 
\textbf{Target(s)} & 
\textbf{Updated CBE $ \;$ allocation} & 
\textbf{Potential changes or simplification(s) $ \; \;$ for repeats} \\
\hline \\

Core throughput (CT) & Yes & $\sim$1.5 & 1 star, V $\geqslant$ 10.7 mag & 2.2\% & Will be defined upon analysis \& repeatability \\ \\

Absolute Flux (AF) & Yes & $\sim$2.75 & 1 star, V$\sim$1.9 mag$\;$  3 stars, V$\sim$12 mag & 3\% & Will be defined upon analysis \& repeatability. For some campaigns we use only 1 faint star to compare with the initial campaign.\\ \\

Astrometric & Yes & $\sim$1.4 & Crowded Field (LMC) & 3\% & Will be defined upon analysis \& repeatability \\ \\

Flat Field (FF) & Yes & $\sim$0.8 & Neptune or Uranus & 0.8\% & Will be assed \\ \\

Image Corrections (IC) Darks \& Clock Induced Charges & Yes & 8.4 & 4 stars, with V mag 4.2, 7.6, 9.7, 11.8 &  1\% &  \\ \\

CTI & Yes & 0 & - & 2.6\%&   \\ \\
Detector $ \; \; \; \; $Backgrounds& Yes & 0 & - & 1.5\%&   \\ \\

\hline \\
Polarimetry & No & $\sim$5.9+1.2 +0.9 & Mueller Matrix and CALSPEC  & TBD\cite{anche2026tmp, mizuki2026tmp} &  Will be defined upon analysis \\
Spectroscopic & No & $\sim$8.4+2.5 +3.25+1.3& IC, CT, Abs Flux, Wavelength Cal.  & TBD &  Will be defined upon analysis \\
\hline
\end{tabular}
\vspace{0.15cm}
\caption{Summary of most of the science calibrations. When the clock time is 0 it means WFI is prime and the \cgi is secondary.}
\label{tab:calplan}
\end{table*}

\subsection{Calibrations Timing Budgets}
\label{sec:timing}

In table~\ref{tab:calplan} we list the same calibration as above but with the corresponding clock time in hours for each item or group of items. We can see that for the HLC NF B1 alone, we are spending already about $\sim$15 hours of "\cpp time" for calibrations even with our slightly looser strategy. It is clear that we will have to assess how stable and how often these calibrations need to be repeated and whether they can be shortened or have to be expanded to achieve our goals. Goal modes like polarimetry and spectroscopy present even more time consuming calibrations but we plan on a lesser cadence.

\section{Pilot Program: a moderate contrast companion}
\label{sec:pilot}  

\noindent The concept of this short program ($\leqslant$ 4 hours, including 2 rolls and a reference star) it to image a relatively bright known stellar companion (flux ratio between  $10^{-6}$ and $10^{-5}$) as early as possible during commissioning. We could thus exercise all our tools including to prepare the observation, simulate the data with a refined orbital solution and astrometric prediction, run the \corgidrp pipeline\cite{wang2026tmp} using simulated calibrations that match the baseline calibration plan described in section~\ref{sec:calplan}. 

\noindent Such observation, if carried out as soon as the \cgi is able to stare at a bright star and dig a preliminary (or "quick") dark hole to the  $\sim$$10^{-7}$ or even $\sim$$10^{-6}$ raw contrast level would ensure that the whole system is working well. I would provide an end-to-end test of the observing pipeline, helping validate image headers, orientation, detector parameters, and data reduction while familiarizing \cpp members and the larger community with the instrument before primary science operations and possibly before attempting the more difficult brown dwarf companion HIP 71618 B\cite{elmorsy2025, elmorsy2026tmp} now foreseen to validate the TTR5 requirements. Based on previous experience with commissioning coronagraphs on-sky, we think it is wise to proceed by small increments in difficulty and learn and adjust along the way.

\subsection{Target selection}
\label{sec:targets}  

Self-luminous companions are ideal because they have well-characterized brightness. White dwarf companions (e.g., HD~114174~B, used to commission the \nircam coronagraphs\cite{girard2022} and whose orbit is well known) are especially attractive due to their well-known, relatively flat spectral energy distributions (SEDs). Brown dwarfs require substantially higher contrast (typically $<10^{-7}$), making them more challenging, while M-dwarf companions offer a good compromise because they are brighter and have well-established visible-light photometry. 

Rather surprisingly, there are not many suitable companions for this activity as we have several constraints:
\bi
\small
\item Candidate systems should have host stars with $V < 6$ (preferably closer to 5). That is because at this point the low order wavefront sensing loop (LOWFS) has not been tested on fainter stars (or equivalent flux in the laboratory).
\item The expected A/B flux ratio should be between $10^{5}$ and $10^{6}$ in Band 1 to make this observation significantly easier and quicker than the TTR5. Ideally no post-processing should be necessary to recover point source B and depending on the available raw contrast and speckle noise floor, PSF subtraction might be optional as described in figure~\ref{fig:dm}. The idea being that we have an immediate assessment of where the companion is.  Also, using such a bright companion enables the operation in "analog" (EMCCD gain $\sim$1) mode which should also be a good functional test.
\item Companion separations should fall approximately in the field of view of the HLC B1 mode: $150\text{--}450~\mathrm{mas}$ (ideally $200\text{--}400~\mathrm{mas}$) during late 2026.
\item The systems should  observable in late 2026/early 2027, preferably in Roman's Continuous Viewing Zone (CVZ with 100\% observability in case anything happens and the schedule changes). 
\ei

\noindent Although HD~114174~B has an ideal predicted 2026.9 separation ($\sim335~\mathrm{mas}$ or $\sim$6$\lambda/D$) and a well-constrained orbit, its host star is slightly too faint (V$\sim$6.4). A literature and archival search was therefore conducted to identify alternative candidate systems using previous high-contrast imaging surveys, commissioning observations, and archival datasets. We curated a long list of self-luminous companions (several tens), some of which could also be used later to test "goal modes" or simply to study some of them once the peak performances are attained or fainter host star operations are possible. 

\subsection{Preparation with the M-dwarf companion to HD 29992}
\label{sec:prep}  

\begin{figure}[h!]
\begin{center}
\includegraphics[width=0.99\textwidth, angle=0]{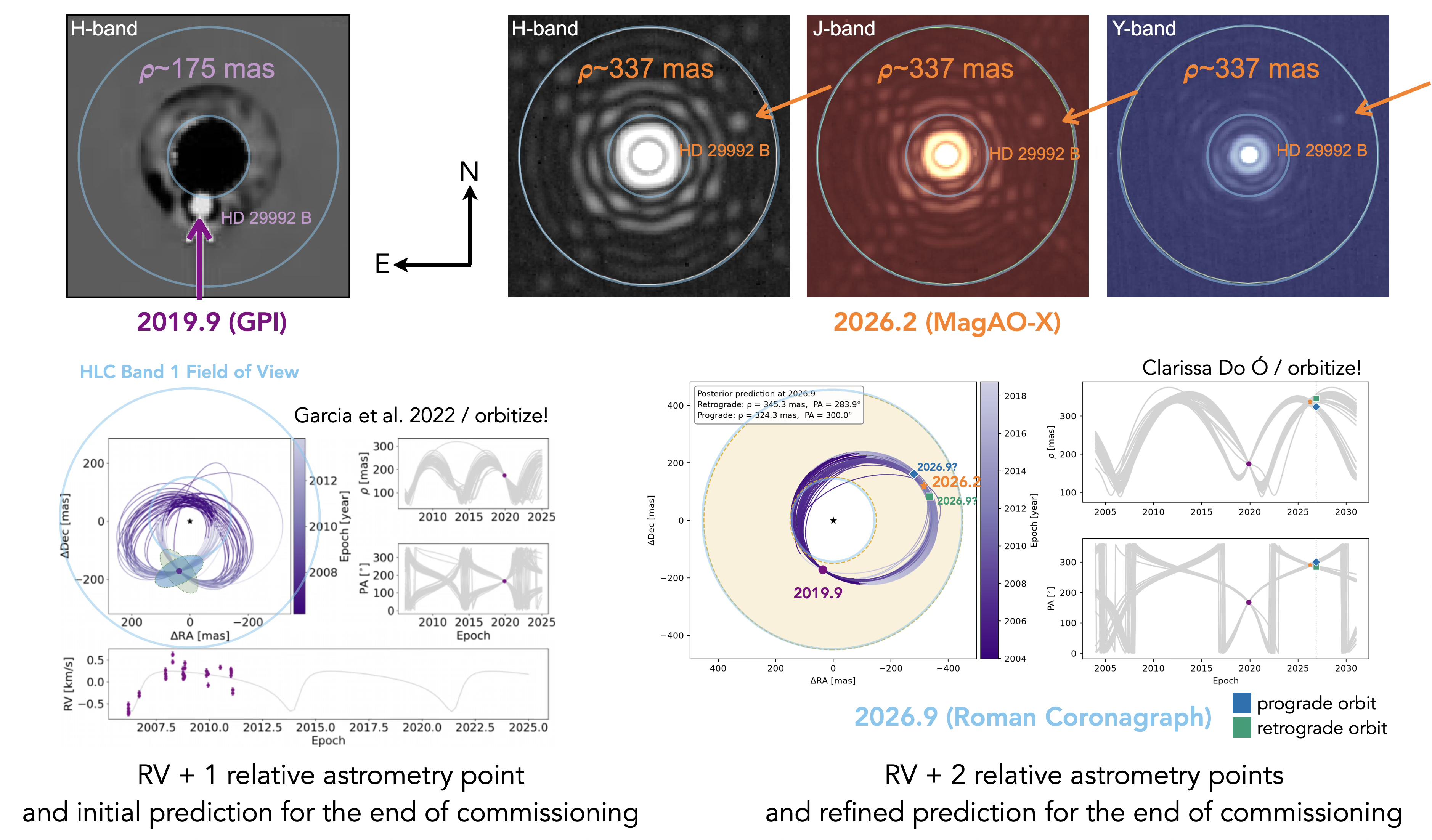}
\vspace{0.2cm}
\caption{HD 29992 B only had been imaged once by GPI in 2019 at a separation of $\sim$175 mas. The orbital fit with this one relative astrometry point combined with radial velocities gave a decent fit estimation of the period ($\sim$ 7.7 years)  but degenerate in inclination (bottom left plot). We reproduced the same plot as in Garcia et al. 2022\cite{garcia2022}. We got a second epoch with MagAO-X in March 2026 and the companion is surprisingly much further away (337 mas) than expected if the period had been correct (green zone roughly at the same position, hence very close to the IWA of the Roman HLC Band 1 mode. Our new fit with \orbitize and two very constraining (because opposite) relative astrometry points predicts a separation $\rho \sim$ 324 mas (if the orbit is prograde) or $\sim$ 345 mas (if the orbit is retrograde) in December 2026 at the end of commissioning (assuming the launch does not slip significantly) and a position angle of $\sim$ 284\arcdeg or $\sim$ 300\arcdeg.}
\label{fig:astrom}
\end{center}
\end{figure}

HD 29992 "ticks all the boxes" (5.1 mag in V, in the CVZ South, the companion to central star B1 flux ratio is estimated to be between  $3\times10^{-5}$ and $3\times10^{-6}$. The prediction of its separation at epoch 2026.9 or 2027.0 turned out to be trickier than anticipated as shown in figure~\ref{fig:astrom}. 

\subsection{Orbital fit refinement}
\label{sec:orbit}  
HD 29992 B was imaged for the first time by the Gemini Planet Imager in 2019\cite{garcia2022}. A relatively good orbital fit was published but with only one imaging relative astrometry measurement, there were two main scenarios (depending on the not well constrained inclination). Nevertheless, trusting that one of the two scenarios was correct and the period was $\sim$7.7 years, we thought the companion would "land" in 2027 very close to the 2019 epoch. Since there was a possibility that the companion would be very close to the IWA, we reached out to the MagAO-X team who had a test run in March and they were able to observe it with a very high Strehl ratio (figure~\ref{fig:astrom}, top right three images) in H, J and Y band. 

\subsection{Photometric check}
\label{sec:photom}  

\noindent Not only these images were extremely useful to refine the orbit, having images from 1 to 1.65$\mu$m allowed us to check the slope of the SED of the M-dwarf companion.
In the discovery paper, they estimate an effective temperature to be $\sim$3,600 K and they give two M9V single stars whose measured spectra match that of  HD 29992B: LHS 2924 (SpeX prim) and LHS 2965 (NIRSPEC BDSS). Using their published photometry, we estimate ("back of the envelope" calculation) the $\Delta_{B1}$ that would apply to HD 29992 B and find the flux ratio range of $3\times10^{-6}$ to $3\times10^{-5}$. The quick and basic photometry performed on the three MagAO-X image seem to favor a higher B/A flux ratio, likely close to $\sim$$\times10^{-5}$ than $\sim$$\times10^{-6}$. The new orbital fit also yielded a higher mass (rather 160\mjup than 80\mjup) and this could confirm it is nowhere near the brown dwarf regime and rather a mid-type red dwarf (to be confirmed with more accurate photometry, and perhaps Roman, soon).

\subsection{Contrast consideration}

\begin{figure}[h!]
\begin{center}
\includegraphics[width=0.8\textwidth, angle=0]{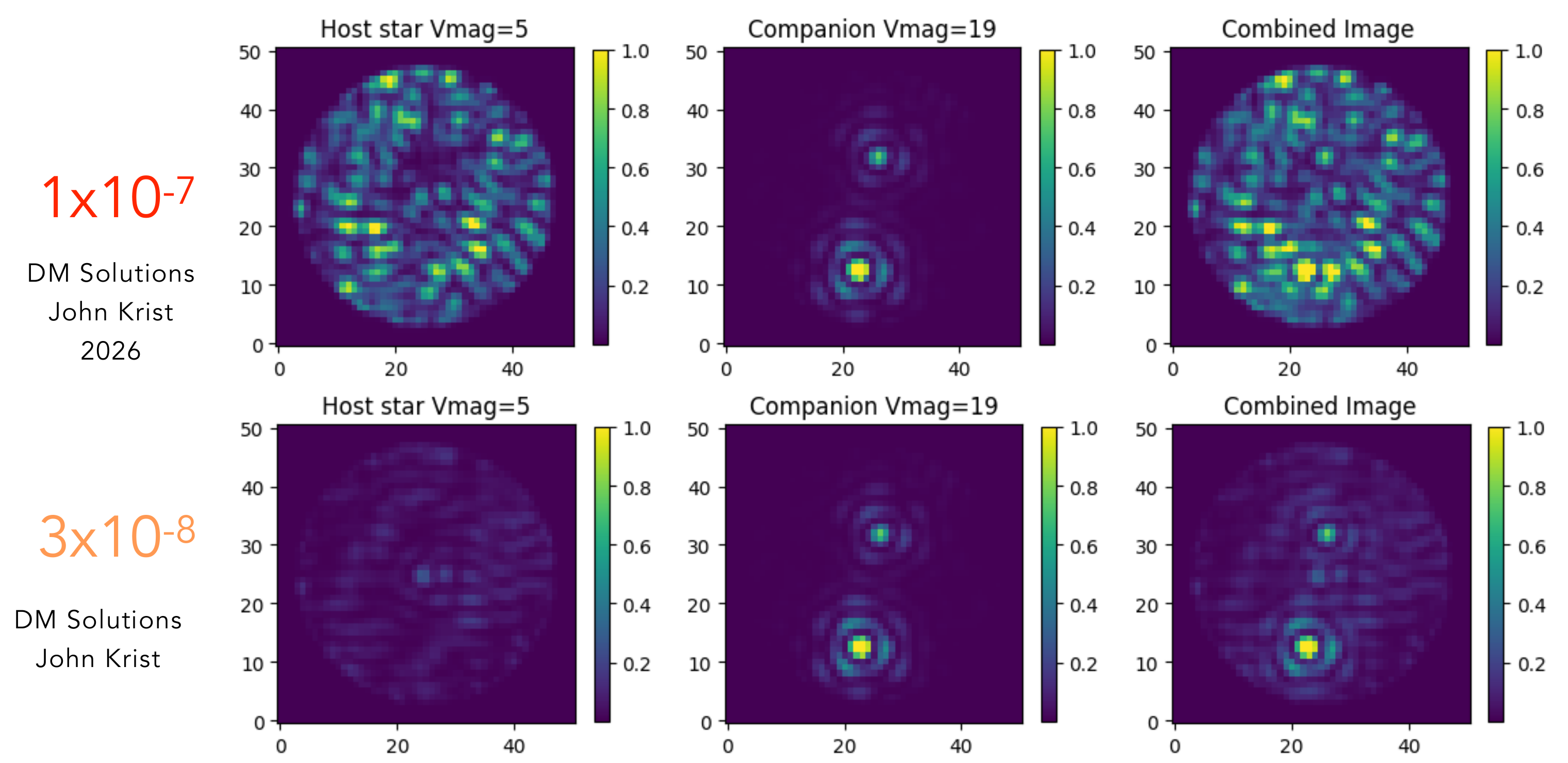}
\vspace{0.2cm}
\caption{Here we used \corgisim to simulate 1.6ks integrations of HD 29992 and check the influence of the core throughput and raw contrast (dark hole depth) with pessimistic deformable mirror (DM) solutions on top and more of a nominal (as it converged during the thermal vacuum tests) residual speckle field at the bottom. The left most panel shows the V=5 mag star behind the coronagraph. The same intensity stretch is used to reveal the difference speckle (photon noise) floor. We injected two $\sim$$10^{-6}$ companions, one at 3$\lambda/D$ (top, like the previously expected separation of HD 29992 B) and one at 6$\lambda/D$ (bottom, similar to the updated expected separation of HD 29992 , figure~\ref{fig:astrom}). The position angles here are irrelevant.}
\label{fig:dm}
\end{center}
\end{figure}

\noindent We prepared a rather conservative plan for HD 29992 B with 1.600s on target (2 rolls) and a 1.8 V mag reference star within 10\arcdeg solar elongation "pitch" angle to hopefully use $*$gam Peg as reference star, which has been well vetted\cite{hom2026} in terms of absence of companion. As shown in figure~\ref{fig:dm}, the $1\times10^{-6}$ companion should be easily recovered at 6$\lambda/D$ (bottom of the 2 panels to the right). With a $3\times10^{-8}$ raw contrast, no PSF subtraction is even needed as the combined image (bottom right) looks nearly the same as the noiseless/speckle-free input scene (middle). For the less optimistic scenario with $1\times10^{-7}$ raw contrast (deformable mirror solutions as computed by J. Krist), PSF subtraction (roll or reference) will enhance the SNR and characterization a lot. But in both cases, post-processing should not be necessary (and/or change the results too much). 

\label{sec:schedule}  

Of course, this is a basic simulation, without realistic detector effects or cosmic rays but the advanced pipeline products should resemble these images if all the calibrations are properly applied.

\section{Conclusions}
\label{sec: concl}  %

\noindent With the \cpp and all the  \rst \cg stakeholders we are ramping up and getting ready for the commissioning and science verification. A lot of our activities are getting fine tuned and are still work in progress. Several contributions in this conference showed more advanced simulations with \corgisim and ran the DRP on full frame images, affected by cosmic rays, etc. Here we decided to focus on the calibration plan strategy and updates as well as the preparation of the pilot observation of HD 29992 B to really "secure" a moderate contrast observation early on.

\noindent This activity will also allow us to learn about the contrast performances with a quick high-order wavefront sensing (HOWFS) setup, appreciate which PSF subtraction strategy works best at this contrast and flux ratio regime. We will check whether a dedicated reference star is necessary or if we can use any bright star previously imaged (or used to dig the dark hole) or a mini-library and a principal component analysis (PCA) "KLIP" model of the PSF\cite{wang2015_pyklip}.
Just like for \jwst, the observation of moderate contrast companion\cite{girard2022, kammerer2022} allowed to flawlessly test everything end-to-end. So much so that when the Early Release Science observations were carried out\cite{carter2023} a couple weeks after, planet HIP 65426 b was recovered within hours in the automatically generated pipeline products. We still have a lot to test for that to be a reality (e.g. have the system work for any number of frames, in photon-counting mode) but the \cpp proved to be efficient at tackling issues. We all hope and will do eveything we can so that sometimes during 2027, we would have imaged at least one mature giant planet in visible reflected light!

\section*{ACKNOWLEDGMENTS}       

\noindent We warmly thank Luciano H. Garcia who promptly shared the \orbitize posterior file of their initial orbital fits with us. We are very thankful to the Las Campanas Observatory staff and MagAO-X team as they observed HD 29992 during one of their test campaigns. We also thank the SPIE Organizing Committee and the Proceedings Coordinators. \\

\noindent This research was carried out in part at the Jet Propulsion Laboratory, California Institute of Technology, under a contract with the National Aeronautics and Space Administration (80NM0018D0004).  A large part of this work was supported by NASA mainly under grants issued through the Roman Community Participation program (CPP). J.\ Girard was supported by NASA under awards 80NSSC24K0097 (PI Millar-Blanchaer) and 80NSSC26K0150 (PI Girard). Portions of this work were supported under award 80NSSC25K0367 (PI Anche). J.J.\ Wang was supported under award 80NSSC24K0087 and by the Sloan Foundation. M. Millar-Blanchaer \ was supported by NASA under award 80NSSC24K0097. S.G. Wolff is supported by NASA under award 80NSSC24K0217. J.\ Gersh-Range was supported under award 80NSSC25K0373. J.\ Hom is supported under award 0NSSC25K0364 (PI Hom). K.\ Ludwick is supported by JPL under Award No. 1652001. C. Do Ó acknowledges the support of the B. Thomas Soifer Fellowship at the California Institute of Technology.  J.\ Ashcraft was supported by NASA through the NASA Hubble Fellowship grant \#HST-HF2-51547.001-A awarded by the Space Telescope Science Institute, which is operated by the Association of Universities for Research in Astronomy. A. Lau and S. Noiret acknowledge the support by the European Union (ERC, ESCAPE, project No. 101044152). Views and opinions expressed are, however, those of the author(s) only and do not necessarily reflect those of the European Union or the European Research Council Executive Agency.\\

\bibliography{corgi} 
\bibliographystyle{spiebib} 

\end{document}